*Brief communications.*

**Low temperature heat capacity of fullerite $C_{60}$ doped with nitrogen.**

A.M. Gurevich[1], A.V. Terekhov[1], D.S. Kondrashev[1], A.V. Dolbin[1], D. Cassidy[2], G.E. Gadd[2], S. Moricca[2], B. Sundqvist[3].

[1] Institute for Low Temperature Physics & Engineering NASU, Kharkov 61103, Ukraine
[2] Australian Nuclear Science & Technology Organisation, Menai, NSW 2234, Australia
[3] Department of Physics, Umea University, SE - 901 87 Umea, Sweden

Electronic address: dolbin@ilt.kharkov.ua

**Abstract**

The heat capacity $C_m$ of polycrystalline fullerite $C_{60}$ doped with nitrogen has been measured in the temperature interval 2 – 13 K. The contributions to heat capacity from translational lattice vibrations (Debye contribution), orientational vibrations of the $C_{60}$ molecules (Einstein contribution) and from the motion of the $N_2$ molecules in the octahedral cavities of the $C_{60}$ lattice have been estimated. However, we could not find (beyond the experimental error limits) any indications of the first – order phase transformation that had been detected earlier in the dilatometric investigation of the orientational $N_2$-$C_{60}$ glass. A possible explanation of this fact is proposed.

The doping of fullerites can affect significantly their properties and thus extend their applications. The impurity effect upon the properties of $C_{60}$ has been studied most extensively. In particular, the effects of some gases (He, $H_2$, $D_2$, Ne, Ar, Kr, Xe, $N_2$) upon the thermal expansion and the structure of $C_{60}$ at low temperatures have been studied in sufficient detail [1, 2, 3]. The most interesting results include in particular the detection of the first – order phase transition stimulated by the gas impurities in the orientational $C_{60}$ glass [2, 3]. Unfortunately, until now the heat capacity of the gas mixtures in fullerites has been left beyond researchers' attention. However, it is interesting to know how the dissolved gases influence the heat capacity of $C_{60}$ or how the first – order phase transition in the glass manifests itself in the behavior of heat capacity. This knowledge is very important too, because it is practically impossible to avoid contamination of $C_{60}$ with air gases. We think that it is reasonable to start investigation of the low temperature heat capacity of the gases dissolved in fullerites with a $N_2$-$C_{60}$ solution because nitrogen is the main constituent of the air. This is the basic objective of this study.

At room temperature $C_{60}$ has a fcc lattice with one octahedral and two tetrahedral interstitial cavities per one $C_{60}$ molecule . The octahedral cavities are sufficiently large in size (4.14 Å [4]) to house molecules of many gas impurities including, for example, $N_2$ molecules with the gas-kinetic diameter $\sigma = 3.7$ Å [5].

The sample was prepared from high-purity (99,99%) $C_{60}$ powder with the average grain size ~ 100 μm (SES, USA). First, the powder was intercalated with $N_2$ and then compacted. The intercalation was performed at the Australian Nuclear Science and Technology Organization (ANSTO, Australia) under the conditions of $P \sim 200$ MPa, $T = 575$°C and $t = 36$ hours. The intercalation technique is described in [6]. The Thermal Gravimetric Analysis results (ANSTO) showed that the octahedral cavities of $C_{60}$ were filled with $N_2$ to practically 100%. The $N_2$-intercalated $C_{60}$ powder was compacted at Umea University, Sweden, by the technique described in [1]. The investigation was started twelve months after preparation of the sample. As a result, according to X-ray analysis, the $N_2$ concentration decreased to 20% [7].

The heat capacity of the $C_{60}$ sample intercalated with $N_2$ was measured by the method of absolute calorimetry in interval 2 – 13 K. The calorimetric cell was a rectangular copper-foil plate with a germanium resistance thermometer and a film heater ($R = 100$ Ohms) fixed on it. The sample with the mass of 0.3215 g was mounted on the free side of the plate covered with a thin layer of vacuum grease. The mass of the calorimetric cell without a sample was 1.081 g. The heat capacity of the cell was measured separately. The measurement error in the heat capacity of the sample was 3-7% (depending on a temperature interval).

The sample was cooled down to liquid helium temperature by cold conduction through the wires (without using helium gas as heat exchanging gas) in vacuum $10^{-2}$ Torr at room temperature and $10^{-6}$ Torr at 4.2 K. The heat capacity was measured in several series (Fig. 1).

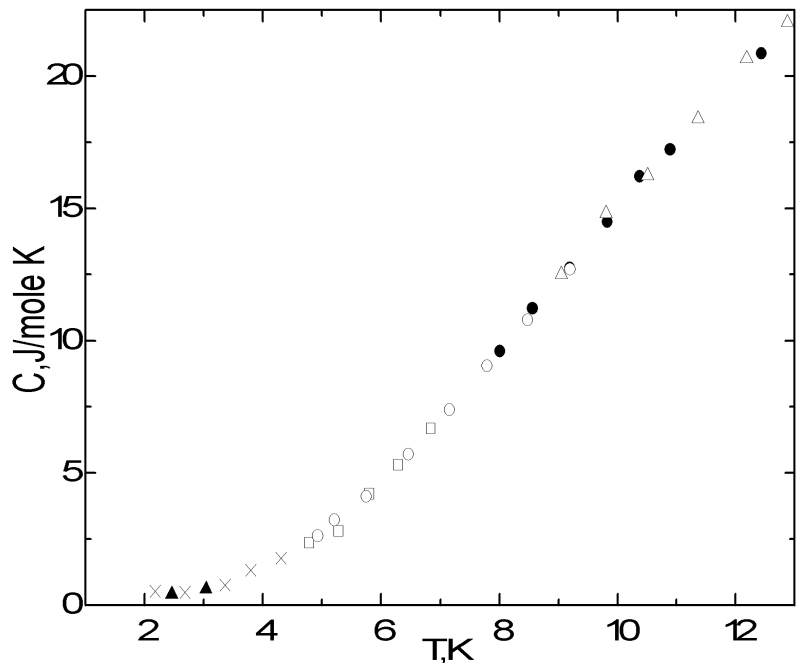

Fig. 1 Molar heat capacity of polycrystalline $C_{60}$ doped with $N_2$.
The measurement series are indicated by different symbols.

The cooling to helium temperature took 24 hours. Then the heat capacity of the calorimetric cell with the sample was measured in a stepwise heating procedure.

We described the temperature dependence of the heat capacity using the following simplified scheme. The heat capacity measured in the interval of experimental temperatures was considered taking into account the contributions from all types of motion: translational fullerite lattice vibrations (Debye contribution), the contribution of orientational vibrations of the $C_{60}$ molecules and the contribution caused by influence of the $N_2$ impurity molecules (Einstein contributions). The experimental results are described fairly well using a Debye term with the characteristic temperature $\Theta_D = 45$ K and two Einstein terms with $\Theta_{E1} = 37$ K and $\Theta_{E2} = 53$ K (Fig. 2). $\Theta_D = 54$K in [1, 8] is the most reliable value for pure $C_{60}$. It is quite reasonable that our $\Theta_D$ is lower because (i) the intercalation with $N_2$ increases the molar volume of the $C_{60}$ crystal and (ii) the added impurity increases the effective molar weight of the system.

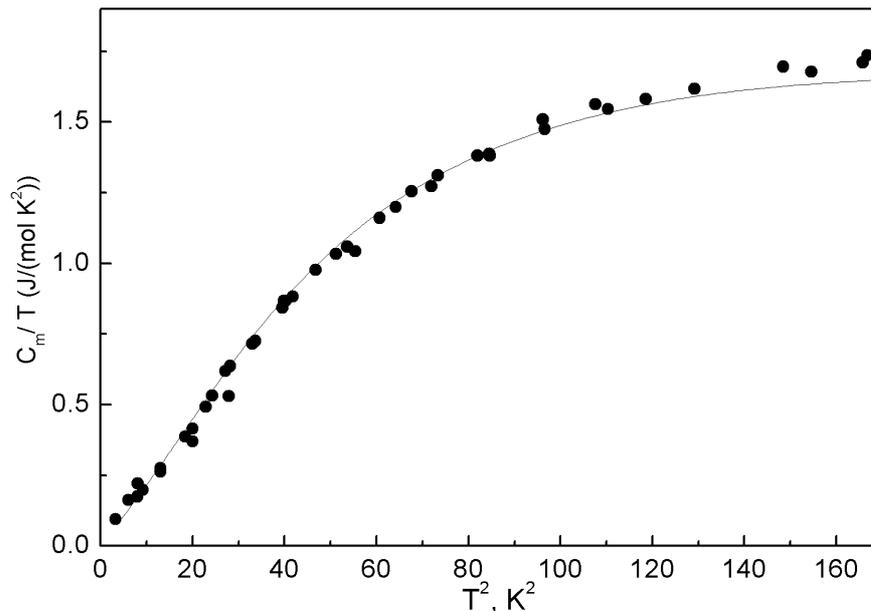

Fig. 2 Calculated and measured temperature dependences of molar heat capacity of polycrystal $C_{60}$ doped with $N_2$:
● – measured $N_2$-$C_{60}$ heat capacity data;
curve – calculated $N_2$-$C_{60}$ heat capacity.

Our model ignores two other contributions to heat capacity. One of them is the linear contribution to the heat capacity of glasses. It can be responsible for the discrepancy between the experimental and calculated data at the lowest temperatures of the experiment. The intramolecular vibrations of $C_{60}$ molecules also contribute to the heat capacity of $C_{60}$. Their contribution may account for the excess of the experimental values over the calculated ones in the high-temperature region of our investigation.

It is known from literature [1, 2, 3, 9] that the temperature dependence of the thermal expansion of $C_{60}$ intercalated with some gases at the temperatures of liquid helium and nitrogen has a hysteresis. Adiabatic calorimetry prohibits measurement of heat capacity at lowering temperature. However, a hysteresis can appear in heat capacity measurement even on heating provided that the thermal prehistory of the sample and the kinetic parameters of the experiment are modified. To detect a hysteresis, the measurement series were made at different starting temperatures and with different rates of heating to the starting temperature. The temperature interval ("step") of measurement was also varied. The sample was cycled at a certain pre-assigned temperature. In this case the heat capacity of the sample was measured at temperature $T_1$. The sample was then cooled down by 1-2 degrees to $T_2$ and the heat capacity was measured at $T_2$. Finally, the sample was heated and the heat capacity was measured at $T_1$ again.

The measured heat capacities are within the limits determined by the experimental error. The data obtained suggest that the heat capacity of the $N_2$-$C_{60}$ system is insensitive to the processes provoking the hysteresis effect. This may be due to the fact that, according to [1, 2], the hysteresis is caused by a first-order phase transition between the orientational glasses based on gas-doped fullerite. However, the difference between the molar volumes of the phases in temperature interval 2-24 K is only 0.02% [1]. In this case the transition-related change in the heat capacity can be smaller than the experimental error. It should be remembered that the phase transition in glasses does not proceed at a constant temperature, it occurs in a rather wide temperature interval. At the same time the phase transformation is clearly indicated by a change in the linear expansion coefficient [1, 2]. The reason may be that that the phase transition considered above is a tunnel transition during which the change in the volume expansion coefficient can be several orders of magnitude larger than the relative change in the heat capacity [10]. It correlates well with the very high Grüniesen coefficients in tunnel energy spectra [11, 12].

To conclude, we emphasize that this study is the first attempt to estimate the effect of gas impurities upon the heat capacity of $C_{60}$ and to find out the possibilities of further research in this direction. Since the molecules of gas impurities have small sizes and masses, we can hardly expect a significant impact of the dissolved gas upon the heat capacity of $C_{60}$ at temperatures far from the phase transition interval. Nevertheless, the increase in the heat capacity of $C_{60}$ at low temperatures observed after introduction of 20% $N_2$ exceeds considerably (by approximately 15%) of the effect that might be caused by the impurity-induced change in the molar volume and the effective mass of $C_{60}$.

Our attempt to reveal the effect of the phase transition on the heat capacity in the orientational glass $N_2$-$C_{60}$ has gone unrewarded.

It is evident that further investigations should be made on solutions with higher concentrations of gas impurities in which the molecules have larger sizes and masses. With these conditions met, both the impurity contribution and the effect produced by the phase transition are expected to increase appreciably. It is also necessary to improve the technique of calorimetric measurement in particular, to ensure the possibility of long-term keeping the solution samples at a constant temperature. We are planning further investigations to pursue this goal.

We wish to thank V.G. Manzhelii and M.I. Bagatskii for helpful participation in the discussion of the results.

**References.**
1. A.N. Aleksandrovskii, A.S Bakai, A.V. Dolbin, V.B. Esel'son, G.E. Gadd, V.G. Gavrilko, V.G. Manzhelii, S. Moricca, B. Sundqvist, B.G. Udovidchenko, *Fiz. Nizk. Temp.* **29**, 432-442, (2003) [*Low Temp. Phys.* **29**, 324-332, (2003)].
2. A.N. Aleksandrovskii, A.S. Bakai, D. Cassidy, A.V. Dolbin, V.B. Esel'son, G.E. Gadd, V.G. Gavrilko, V.G. Manzhelii, S. Moricca, B. Sundqvist, *Fiz. Nizk. Temp.* **31**, 565, (2005) [*Low Temp. Phys*. **31**, 429 (2005)].
3. V.G. Manzhelii, A.V. Dolbin, V.B. Esel`son, V.G. Gavrilko, D. Cassidy, G.E. Gadd, S. Moricca, and B. Sundqvist, *Fiz. Nizk. Temp.* **(**to be published**)**.
4. M.S. Dresselhaus, G. Dresselhaus, and P.C. Eklund, Science of Fullerenes and Carbon Nanotubes, Academic Press, San Diego, California, 1996.
5. V.G. Manzhelii, M.A. Strzhemechny, Y.A. Freiman, A.I. Erenburg, V.A. Slusarev, Physics of Cryocrystals, AIP Press, American Institute of Physics, Woodbury, New Yourk (1997).
6. G.E. Gadd, S. Moricca, S.J. Kennedy, M.M. Elcombe, P.J. Evans, M. Blackford, D. Cassidy, C.J. Howard, P. Prasad, J.V. Hanna, A. Burchwood and D. Levi, *J. Phys. Chem. Solids* **56**, 1823 (1997).
7. A.I. Prokhvatilov, private communication.
8. N.A. Aksenova, A.P. Isakina, A.I. Prokhvatilov, and M.A. Strzhemechny, *Fiz. Nizk. Temp.* **25**, 964 (1999) [*Low Temp. Phys.* **25**, 724, (1999)].
9. A.I. Prokhvatilov, N.N. Galtsov, I.V. Legchenkova, M.A. Strzhemechny, D. Cassidy, G.E. Gadd, S. Moricca, B. Sundqvist, N.A. Aksenova, *Fiz. Nizk. Temp.* **31**, 585 (2005) [*Low Temp. Phys*. **31**, 445 (2005)].
10. Y.A. Freiman, *Fiz. Nizk. Temp.* **9**, 567 (1983).
11. C.R. Case, K.O. McLean, C.A. Swenson, and G.K. White, Thermal Expansion-1971, AIP Conference Proc., New York (1972), p. 312.
12. C.R. Case and C.A. Swenson, *Phys. Rev. B* **9**, 4506 (1974).